\documentclass[11pt,nofootinbib,floatfix,onecolumn,preprintnumbers]{revtex4}
\pdfoutput=1

\usepackage{amsmath,amssymb}
\usepackage{graphicx}
\usepackage{rotating}
\usepackage{color}
\usepackage{xcolor}
\usepackage{multirow}
\usepackage{fontenc}
\usepackage{slashed}
\usepackage{longtable}
\usepackage{dcolumn}
\usepackage{bm}
\usepackage{epsfig}
\usepackage[colorlinks,bookmarksopen,bookmarksnumbered,linkcolor=blue,pdfstartview=FitH,urlcolor=blue,citecolor=blue]{hyperref}

\newcommand{\be}{\begin{equation}}
\newcommand{\ee}{\end{equation}}
\newcommand{\ba}{\begin{eqnarray}}
\newcommand{\ea}{\end{eqnarray}}
\newcommand{\bea}{\begin{eqnarray*}}
\newcommand{\eea}{\end{eqnarray*}}
\newcommand{\nn}{\nonumber}

\newcommand{\AddrKim}{%
Department of Physics and IPAP, Yonsei University, Seoul 120-749, Korea
}

\newcommand{\AddrGabriel}{%
Departamento de F\'isica, Centro de Investigaci\'on y de Estudios Avanzados,
Apartado Postal 14-740, 07000 M\'exico Distrito Federal, M\'exico
}

\begin{document}

\title{Evaluation of  CKM matrix elements from  exclusive $P_{\ell 4}$ decays}

\author{C. S. Kim}\email{cskim@yonsei.ac.kr}
\affiliation{\AddrKim}

\author{G. L\'opez Castro}\email{glopez@fis.cinvestav.mx}
\affiliation{\AddrGabriel}

\author{S. L. Tostado}\email{stostado@fis.cinvestav.mx}
\affiliation{\AddrGabriel}

\begin{abstract}
\noindent We consider the exclusive $P_{\ell 4}$ decays, $P\to (P_1P_2)_V\ell \nu_{\ell}$,
where the subindex $V$ means that the invariant mass of the pseudoscalar pair is taken within a small window around the mass of the vector  meson $V$.  Pole contributions beyond the dominant $P\to V(\to P_1P_2)\ell\nu_{\ell}$ amplitude of $P_{\ell 4}$ decays are identified, which, in turn, affects the determination of the CKM matrix elements $|V_{qq'}|$. We evaluate the effects of those contributions in the extraction of  bottom and charm quark mixings. An application to $B\to (\pi\pi)_{\rho}\ell \nu_{\ell}$ data from Belle collaboration, shows an increase in the extracted value of $|V_{ub}|$ in better agreement with determinations based on $B\to \pi\ell\nu_{\ell}$ decays.
The effect of the $\rho$ and $D^*$ pole contributions in the determination of $|V_{cd}|$ from the decay 
$D\to \pi\pi \ell^- \bar{\nu}_{\ell}$, has been also investigated.
\end{abstract}

\maketitle

\section{Introduction}
\label{sec:intro}

Precise measurements of the Cabibbo-Kobayashi-Maskawa (CKM)   \cite{Cabibbo:1963yz,Kobayashi:1973fv} matrix elements can shed light on new physics and are among the main targets of flavor factories. Deviations from the unitarity property of the CKM matrix would indicate the existence of additional degrees of freedom. While the determination of $|V_{ud}|$ and $|V_{us}|$ has been done with an impressive accuracy of 0.02\% and 0.3\%, respectively,  the values of  $|V_{ub}|$ and $|V_{cb}|$ are known at the 5\% and 2\% only \cite{Olive:2016xmw}.  Better determinations of  $|V_{qb}|$, among other standard model parameters, are   the most important for searches of new sources of CP violation beyond the one encoded in the CKM paradigm.


Currently, the most precise determinations \cite{Olive:2016xmw} of the matrix elements, $|V_{cb}|$ and $|V_{ub}|$, 
indicate a tension between values extracted from  exclusive and inclusive decay channels of 
$b$-flavored hadrons.
Clearly,  more theoretical works and refined measurements are required to solve this discrepancy and achieve a better accuracy. One can gain some precision  by combining values of $|V_{qb}|$ extracted from different decay channels of bottom hadrons, provided their measurements furnish a consistent set of data. While measurements of exclusive channels are better suited from an experimental point of view, the calculation of their form factors  in the whole kinematical regime is still challenging. Among the preferred exclusive channels, the $B\to (P,V)\ell \nu_{\ell}$ decays, with $P$ ($V$) a pseudoscalar (vector) meson, are the simplest ones to describe theoretically and the dominant final states of charmfull and charmless semileptonic decays of $B$ mesons.  While Lattice QCD provides reliable results at low recoil of final state mesons \cite{Bernard:2008dn,Na:2015kha,Lattice:2015rga,Lattice:2015tia}, other methods  (like Light Cone sum rules, see for example \cite{Khodjamirian:2005ea,Ball:2004rg,Ball:2004ye}) are better suited at larger recoil values. Finally, experimental data can be used as a guide to extrapolate between these two domains.

 In this paper we are concerned with the extraction of $|V_{qq'}|$ matrix elements from $B,D \to V\ell \nu_{\ell}$ and its related observable $B,D \to P_1P_2\ell\nu_{\ell}$ decay channel. As it was mentioned above, these exclusive decays provide complementary information on CKM matrix elements and a consistency test of  values extracted from other exclusive and inclusive channels. Furthermore, a good understanding of the dominant  exclusive channels is essential to describe how inclusive decays are built out from exclusive components.

While pseudoscalar mesons are quasistable states, some of them directly detectable by experiments, vector mesons are highly unstable resonances, which are reconstructed from their detectable decay products. From the theoretical point of view, using vector mesons as asymptotic states of the $S$-matrix is an approximation which, in principle, is not justified owing to their very short lifetimes. A theoretical definition, that is consistent with the experimental one, can be used instead.   In this paper we will consider $P\to P_1P_2\ell\nu_{\ell}$ ($P_{\ell 4}$) transitions, where the $P_1P_2$ pseudoscalar pair is produced {\it dominantly} from a decay of a single vector meson  $V\to P_1P_2$.  The extraction of the decay observables associated to $P\to V\ell \nu_{\ell}$ decays is affected by the contributions of subdominant $s$ and $d$ wave configurations of the $P_1P_2$ system \cite{Faller:2013dwa, Kang:2013jaa, Meissner:2013pba, Albertus:2014xwa, Hambrock:2015aor, Cheng:2017smj, Shi:2017pgh} even if one chooses a narrow window in their invariant mass distribution around the $V$ resonance mass.
 For an example, the extraction of the $|V_{ub}|$ matrix element from $B\to \pi\pi\ell\nu_{\ell}$ considering the resonances, backgrounds and rescattering effects in the $\pi\pi$ system, were studied  in Refs. \cite{Kang:2013jaa, Meissner:2013pba, Albertus:2014xwa, Straub:2015ica}. Those authors found that these effects can bring the determination of $|V_{ub}|$ from four-body semileptonic decays in better agreement with the value extracted from $B\to \pi\ell \nu_{\ell}$ decay.
Also, additional kinematical distributions accessible in four-body semileptonic decays, as compared to three-body decays, allows to explore further observables sensitive to new physics \cite{Meissner:2013pba}.  A study of $D^+\to K^-\pi^+ e^+ \nu_{e}$ decays that incorporate the strong interaction dynamics of the $K\pi$ system was recently reported in Ref. \cite{Ablikim:2015mjo}.

Here, we consider the effects of an additional pole contribution $P^*$ in the observables associated to  $P\to P_2P^*\to P_1P_2\ell\nu_{\ell}$ decays. Although four-body decays of heavy mesons have been considered before including refinements in the treatment of the $s$-wave and excited resonances in the $p$-wave of final state mesons \cite{Faller:2013dwa,Meissner:2013pba,Kang:2013jaa,Albertus:2014xwa,Straub:2015ica,Hambrock:2015aor,Cheng:2017smj, Shi:2017pgh}, the effects of the $P^*$ pole has not been considered in the literature. This pollution can affect the different invariant mass distribution of the $P_1P_2$ system and can modify the values of CKM matrix elements extracted from $P\to V \ell \nu_{\ell}$ transitions. Examples of these decays are $B\to (D\pi, \pi\pi) \ell\nu$ or $D\to (K\pi, \pi\pi)\ell \nu$ which are dominated by the $(D^*,\rho)$ and $(K^*, \rho)$ resonances, respectively. The presence of additional $B^*$ and $D^*$ poles can affect the determination of the  CKM matrix elements to a few percent level, which are important for present and future studies. We present the effects of these additional pole contributions in the invariant mass distribution of the meson pair and in the branching fractions,
 to estimate their effects in the values of the relevant CKM matrix elements.

\section{Four-body semileptonic decays of pseudoscalar mesons}
\label{sec:BtoDstar}

  Let us consider the generic $P(p)\to P_1(p_1)P_2(p_2)\ell(p_3)\nu_{\ell}(p_4)$ decay, denoted as $P_{\ell 4}(P_1P_2)$, induced by the quark level transition $q\to q'\ell \nu_{\ell}$, with $(p,p_i)$ the particle four-momenta subject to the on-shell conditions ($p^2=M^2,\ p_i^2=m_i^2$). At the lowest level, using the local approximation (infinitely heavy $W$ boson), the decay amplitude can be written as
\be
{\cal M} = \frac{G_F}{\sqrt{2}} V_{qq'} H_{\mu}\ell^{\mu}\ ,
\ee
where $V_{qq'}$ is the quark mixing CKM matrix element,  $\ell_{\mu}$ is the leptonic $V-A$ charged current, and
\ba
H_{\mu}&=& \langle P_1(p_1)P_2(p_2) | j_{\mu}| P(p) \rangle \ \nn \\
&=& V_{\mu}-A_{\mu}
\ea
is the hadronic matrix element of the $V-A$ quark current.

 Following Ref. \cite{Bijnens:1994me}, we can write the most general  vector and axial-vector pieces of the hadronic matrix element as follows:
\ba
V_{\mu}&=&-\frac{H}{M^3} \epsilon_{\mu\nu\rho\sigma}L^{\nu}P^{\rho}Q^{\sigma}\ ,\label{vff} \\
A_{\mu} &=& -\frac{i}{M} \left[F P_{\mu} +GQ_{\mu} +RL_{\mu} \right] \ .  \label{aff}
\ea
The form factors $H, F, G, R$ depend on the square of the momentum transfer to leptons 
and on two additional independent Lorentz scalars \cite{Cabibbo:1965zzb,Bijnens:1994me}.
The hadronic vertices (\ref{vff},\ref{aff}) depend upon  three-independent Lorentz vectors which we chose as $P=p_1+p_2,\ Q=p_1-p_2,\ L=p_3+p_4=p-p_1-p_2$. Conservation of energy-momentum implies $p=P+L$. This choice is useful to fix the set of  five independent kinematical variables to describe the four-body decay: $(s_{12}=P^2,\ s_{34}=L^2,\ \theta_{P},\ \theta_{\ell}, \phi)$ (see definitions in Refs. \cite{Cabibbo:1965zzb,Bijnens:1994me}). The corresponding limits of integration are given by (for massless neutrinos $m_4=0$): $(m_1+m_2)^2\leq s_{12}\leq (M-m_3)^2;\ m_3^2\leq s_{34} \leq (M-\sqrt{s_{12}})^2;\ 0\leq \theta_P,\ \theta_{\ell}\leq \pi$ and $0\leq \phi \leq 2\pi$ \cite{Cabibbo:1965zzb,Bijnens:1994me}.

  One can get the decay rates by integrating over $s_{12}$, the invariant mass distribution of the pair of final state  pseudoscalar mesons
\be
\Gamma(P\to P_1P_2\ell\nu_{\ell})=\int_{s_{12}^-}^{s_{12}^+} ds_{12} \frac{d\Gamma(P\to P_1P_2\ell\nu_{\ell})}{ds_{12}}\ . \label{integratedrate}
\ee
In the case that the invariant mass distribution is fully dominated by a single intermediate resonance $R$, namely $P\to R(\to P_1P_2)\ell\nu_{\ell}$, we can restrict the integration to the region defined by $s_{12}^{\pm}=(m_R\pm \Delta)^2$, where $m_{R}$ is the mass of the resonance, and typically $\Delta=\Gamma_R/2$ or $\Gamma_R$, with $\Gamma_R$ its decay width. In the case of a very narrow resonance ($\Gamma_R\to 0$), one recovers the usual result
\be
\Gamma(P\to P_1P_2\ell\nu_{\ell})= \Gamma(P\to R\ell\nu_{\ell}) \times B(R\to P_1P_2)\ .
\ee
This result is also a good approximation for wider resonances, provided no other contributions to the decay amplitude are present. It will be modified, however, by the contribution of additional pole contributions to the decay amplitude.

\begin{figure}\centering
\includegraphics[width=14.0cm]{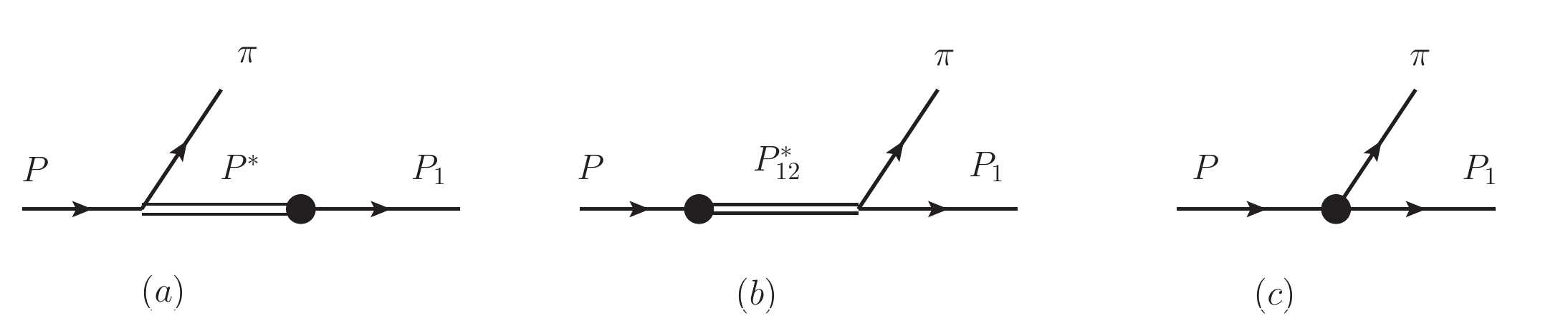}
  \caption{\small Contributions to the hadronic vertex in $P\to P_1\pi \ell^-\bar{\nu}_{\ell}$ decays. Double-lines are used for the intermediate vector resonances.  The solid dot indicates the hadronic weak vertex. }\label{figu1}
\end{figure}

As shown in Figure \ref{figu1}, there are three different contributions to the hadronic vertex of the $P\to P_1P_2\ell \nu$ decay (here we have chosen $P_2=\pi$, for definiteness). We will assume that the two dominant contributions are given by single pole contributions, namely: $P\to P^*\pi  \to P_1\pi\ell \nu$ (Figure \ref{figu1}a) and $P\to P_{12}^*\ell \nu\to P_1\pi \ell \nu$ (Figure \ref{figu1}b). Additional resonances in these channels can contribute as well and their contributions can be trivially added to our results; notice that usually the effects of heavier resonances similar to the one of interest in the $P_{12}^*$ channel are taken into account and estimated as background in simulations. For simplicity, we will make the reasonable assumption that pole contributions are dominated by the exchange of vector meson resonances in Figures \ref{figu1}a,b.

 The different hadronic  vertices that enter in decays of charged and neutral $P$ mesons in Figure \ref{figu1}, are related by isospin symmetry which will be assumed as a good approximation.
We define the strong  $PVP'$ vertex as $ig_{_{PVP'}}(p-p')\cdot \epsilon(p_V)$, using the convention $V(p_V,\epsilon_V)\to P(p)P'(p')$.  The weak matrix elements required for our evaluations corresponding to the two vector pole contributions in Figure \ref{figu1} are (we use the convention of Ref. \cite{Chang:2016cdi} for the $R'\to P_1$ transition)
\ba
\langle P_1(p_1)|j_{\mu}|R'(\varepsilon,p_R') \rangle &=& -\frac{2iV'}{m_{R'}+m_1}\epsilon_{\mu\nu\rho\sigma}\varepsilon^{\nu}p_1^{\rho}p_{R'}^{\sigma}-2m_{R'}A'_0 \frac{q'\cdot \varepsilon}{q'^2} q'_{\mu} \nn \\
&& -(m_{R'}+m_1)A'_1 \varepsilon^{\beta}T_{\beta\mu}(q')-A'_2\frac{\varepsilon\cdot q'}{m_{R'}+m_1} (p_{R'}+p_1)^{\beta}T_{\beta\mu}(q')
\ea
and
\ba
\langle R(\varepsilon^*,p_R)|j_{\mu}|P(p) \rangle &=& \frac{2iV}{m_R+M}\epsilon_{\mu\nu\rho\sigma}\varepsilon^{*\nu}p^{\rho}p_R^{\sigma}-2m_RA_0 \frac{q\cdot \varepsilon^*}{q^2} q_{\mu} \nn \\
&& -(m_R+M)A_1 \varepsilon^{*\beta}T_{\beta\mu}(q)+A_2\frac{\varepsilon^*\cdot q}{m_R+M} (p+p_R)^{\beta}T_{\beta\mu}(q)
\ea
for the $P\to R$ transition \cite{Chang:2016cdi}.
The weak current is $j_{\mu}=\bar{q'}\gamma_{\mu}(1-\gamma_5)q$; we have assumed that the intermediate resonances $R\ (R')$ are vector mesons, with polarization four-vector $\varepsilon^*_{\mu}\ (\varepsilon_{\mu})$, such that $p_R\cdot \varepsilon^*=p_{R'}\cdot \varepsilon=0$. The primed form factors for the $R'\to P_1$ weak transition depend upon $q'^2=(p_{R'}-p_1)^2$, while those of $P\to R$ depend upon $q^2=(p-p_R)^2$; owing to energy-momentum conservation $q=q'=L$. In the above expressions  $T_{\beta\mu}(q)\equiv g_{\beta\mu}-q_{\beta}q_{\mu}/q^2$ is the hadronic transverse tensor ($q^{\beta}T_{\beta\mu}=0$).

As a concrete example, let us consider the $B(p)\to D(p_1)\pi(p_2) \ell(p_3) \nu_{\ell} (p_4)$ decay, which we assume to be dominated by the $B^*$ and $D^*$ pole contributions in the region where the $D\pi$ invariant mass is close to the $D^*$ resonance. Using the definitions introduced previously, we can compute the form factors defined in Eqs. (\ref{vff})-(\ref{aff}) and get the following results:
\ba
-\frac{H}{M_B^3} &=&2i\left[-\frac{g_{_{BB^*\pi}}}{D_{B^*}(p-p_2)} \frac{V'}{m_{B^*}+m_D}+ \frac{g_{_{DD^*\pi}}}{D_{D^*}(P)} \frac{V}{M_B+m_{D^*}}\right] \ ,\\
-\frac{iF}{M_B} &=& \frac{g_{_{BB^*\pi}}}{D_{B^*}(p-p_2)} \left[\frac{A'_2X}{m_{B^*}+m_D}+(m_{B^*}+m_D)A'_1X_1 \right] \nn \\
&&-\frac{g_{_{DD^*\pi}}}{D_{D^*}(P)} \left[\frac{2A_2Y}{M_B+m_{D^*}}+(M_B+m_{D^*})A_1\frac{P\cdot Q}{m_{D^*}^2} \right] \ ,\\
-\frac{iG}{M_B}&=&\frac{g_{_{BB^*\pi}}}{D_{B^*}(p-p_2)} \left[\frac{A'_2X}{m_{B^*}+m_D}+(m_{B^*}+m_D)A'_1X_2 \right] \nn \\
&&+\frac{g_{_{DD^*\pi}}}{D_{D^*}(P)}(M_B+m_{D^*})A_1  \ , \\
-\frac{iR}{M_B}&=&\frac{g_{_{BB^*\pi}}}{D_{B^*}(p-p_2)} \left[2m_{B^*}\frac{A'_0X}{L^2}-\frac{(P+Q)\cdot L}{L^2}\frac{A'_2X}{m_ {B^*}+m_D} +(m_{B^*}+m_D)A'_1\left(A-\frac{X\cdot L}{L^2} \right)\right]\nn  \\
&&+\frac{g_{_{DD^*\pi}}}{D_{D^*}(P)}\left[ 2m_{D^*}\frac{A_0Y}{L^2}+\frac{2A_ 2Y}{M_B+m_{D^*}}  \frac{P\cdot L}{L^2} -(M_B+m_{D^*})A_1\frac{Y\cdot L}{L^2} \right] \ .
\ea
In the above expressions we have defined the four-vectors $X^{\beta}=X_+P^{\beta}+X_-Q^{\beta}+AL^{\beta}$, $Y^{\beta}=Q^{\beta}-(P\cdot Q/m_{D^*}^2)P^{\beta}$, and the Lorentz scalars $X=L\cdot X,\ Y=L\cdot Y, \ X_{\pm}=A/2\pm 1$, with $A=1-p_{B^*}^2/m_{B^*}^2[1+(P-Q)\cdot p_{B^*}/p_{B^*}^2]$, where $p_{B^*}=P+L-(P-Q)/2$. We have used the notation $D_{B^*}(q)=q^2-m_{B^*}^2$ and $D_{D^*}(q)=q^2-m_{D^*}^2+im_{D^*}\Gamma_{D^*}$. It is easy to check that for $P\to P_1 \pi \ell \nu_{\ell}$ decays we have to replace $B^{(*)}\to P^{(*)}$, $D^{(*)}\to P_1(P_{12}^*)$ and the corresponding weak form factors and strong coupling constants in the previous expressions.

\section{Discussions on Experimental Observables of $P_{\ell 4}$}

Several experiments have reported measurements of branching ratios and invariant mass distributions of $B_{\ell 4}$ \cite{Behrens:1999vv,delAmoSanchez:2010af,Sibidanov:2013rkk} and $D_{\ell 4}$ \cite{delAmoSanchez:2010fd,CLEO:2011ab,Ablikim:2015mjo} decays.
The corresponding analysis to extract the CKM matrix elements from the $D,B\to V\ell \nu_{\ell}$ observables differ in several ways: $(a)$ the form factors used to model the weak transition, $(b)$ the window of the hadronic mass distribution $|\sqrt{s_{12}}-m_V|$ chosen to isolate the $V$ vector meson signal, and $(c)$ the inclusion of several wave configurations and resonance contributions in the hadronic system. As an illustration of the second item, the following cuts in the hadronic invariant mass distribution have been used by different experiments for  $B\to \pi\pi\ell \nu_{\ell}$ decays: $|\sqrt{s_{\pi\pi}}-m_{\rho}| \leq \Gamma_{\rho}$ \cite{Behrens:1999vv} ( $2\Gamma_{\rho}$ \cite{Sibidanov:2013rkk}),
$0.650\ {\rm GeV} \leq \sqrt{s_{\pi\pi}} \leq 0.850\ {\rm GeV}$ \cite{delAmoSanchez:2010af}, and $0.60\ {\rm GeV} \leq \sqrt{s_{\pi\pi}} \leq 1.00\ {\rm GeV}$ \cite{Hokuue:2006nr}, which prevent a direct comparison of reported values for the branching fractions.
In addition, some experiments report values of the combined results from neutral and charged $B$ meson branching fractions using isospin symmetry. Isospin symmetry breaking effects should be duly taken into account in analyses when measurements reach the one percent accuracy. Furthermore, the contributions of the additional pole contribution considered in this paper become relevant at the few percent level accuracy determinations of CKM matrix elements.

  In our previous paper \cite{Kim:2016yth}, we have shown that the effects of the $B^*$ pole contribution in $B\to D\pi\ell\nu_{\ell}$ is negligibly small compared to the $D^*$ pole, owing to the very narrow width of the $D^*$ resonance which fully dominates the $D\pi$ invariant mass distribution close to the $D^*$ mass. The effect of the $B^*$ pole in the extraction of the ratio $R(D^*)$ from $B_{\ell 4}(D\pi)$ decays is also negligible \cite{Kim:2016yth}.
   This leads to the interesting question of how large this effect can be for wider resonances and how it affects the extraction of the CKM matrix elements when using $P\to V \ell \nu$ decays. Here we study the effects of the pole diagram of Figure \ref{figu1}(a) in the hadronic invariant mass distribution and the branching fraction of $P_{\ell 4}$ decays in the region close to the $P_{12}$ resonance and its consequences for the extraction of $|V_{qq'}|$.

  In our calculations we use the following  phase convention for pseudoscalar meson states \cite{Gibson:1976wp}:  $|\pi^+\rangle =-u\bar{d}, \ |\pi^0\rangle =\frac{1}{\sqrt{2}}(u\bar{u}-d\bar{d}), \ |\pi^-\rangle =d\bar{u}, \ |K^+\rangle =u\bar{s}, \ |K^0\rangle =d\bar{s},\ |\overline{K}^0\rangle =-s\bar{d}, \ |K^-\rangle =s\bar{u}, \  |D^+\rangle =-c\bar{d}, \ |D^0\rangle =c\bar{u},\ |\overline{D}^0\rangle = u\bar{c}, |D^-\rangle =d\bar{c}$. The convention for $B$ mesons are similar to $K$ mesons under the replacement $s\to b$. With these conventions, isospin symmetry provides the following relations among different couplings:
\begin{eqnarray}
g_{_{\rho \pi \pi}} &=& g_{_{\rho^+ \pi^+ \pi^0}} = -g_{_{\rho^0 \pi^+ \pi^-}} ~, \nonumber \\
g_{_{K^{*} K \pi}} &=& g_{_{K^{*+} K^0 \pi^+}} = \sqrt{2}g_{_{K^{*+} K^+ \pi^0}} =-g_{_{K^{*0}K^+ \pi^-}} = \sqrt{2} g_{_{K^{*0}K^0 \pi^0}} ~, \nonumber \\
g_{_{D^* D \pi} } &=& g_{_{D^{*+}D^0\pi^+}}=-\sqrt{2}g_{_{D^{*+}D^+\pi^0}}=-g_{_{D^- \bar D^{*0} \pi^-}}=-\sqrt{2}g_{_{\bar D^0 \bar D^{*0} \pi^0}} ~, \nonumber \\
g_{_{B^* B \pi}} &=& -g_{_{B^- \bar B^{*0} \pi^-}} = -\sqrt{2} g_{_{\bar B^0 \bar B^{*0} \pi^0}}\ .
\end{eqnarray}
For our numerical evaluations we will use the values $g_{_{\rho \pi \pi}}=5.98\pm0.03$, $g_{_{K^{*} K \pi}}=3.28\pm0.03$ and $g_{_{D^* D \pi }}=8.39\pm0.08$ extracted from the experimental widths of resonances \cite{Olive:2016xmw}, and $g_{_{B^* B \pi}}=20.0\pm1.2$ from the most recent lattice calculations \cite{Bernardoni:2014kla}. The masses and widths of the vector resonances are taken from Ref. \cite{Olive:2016xmw}.

  For the weak form factors we use the following results: $(i)$ for the $B\to \rho$ transitions we rely on Lattice calculations of Ref. \cite{DelDebbio:1997ite}; $(ii)$ the form factors for semileptonic decays of charmed mesons  $D\to K^*$  and $D\to \rho$  are taken from experimental data of Refs. \cite{Ablikim:2015mjo} and \cite{CLEO:2011ab}, respectively; $(iii)$ for the evaluation of the $P^*\to P_1$ form factors, we use the relativistic harmonic oscillator potential model of Refs.  \cite{Wirbel:1985ji,Bauer:1988fx} (WSB). In this model, the $q^2$ dependence of all the form factors are assumed to have a monopolar form:
\be
F_i(q^2)=\frac{F_i(0)}{1-q^2/m_i^2(J^P)} \ .
\ee
The form factors at $q^2=0$ are computed from the overlap of relativistic wave functions in this model; the values of pole masses are chosen to correspond to the lightest resonances with appropriate quantum numbers that allows coupling to the weak currents.

\begin{table}\centering
\begin{tabular}{|c|c|c|c|c||c|c|c|}
\hline
Transition & $\ A'_0(0)$\ & \ $A'_1(0)$\  & \ $A'_2(0)$\ & \ $V'(0)$\ & \ $m(0^-)$\  & \ $m(1^-)$\  &\  $m(1^+)$\  \\
\hline \hline
$D^* \to K$ & 0.78 & 1.02 & 0.40  & 0.90 & 1.97 & 2.11 & 2.53 \\
\hline
$D^* \to \pi$ & 0.75 & 1.08 & 0.37 & 0.76 & 1.87 & 2.01 & 2.42 \\
\hline
$B^* \to \pi$  \cite{Chang:2016cdi}& 0.34 & 0.38 & 0.29  & 0.34 & 5.27 & 5.32 & 5.71 \\
\hline
$B^*\to D$  \cite{Chang:2016cdi} & 0.63 & 0.66 & 0.56 & 0.70 & 6.30 & 6.34 & 6.73 \\
\hline
\end{tabular}
\caption{Form factors of the weak transition $P^* \to P_1 $ at $q^2=0$ in the Wirbel-Stech-Bauer model \cite{Wirbel:1985ji,Bauer:1988fx}. Values of pole masses are given in GeV units.} \label{table2}
\end{table}

In the case of $B^* \to D$ and $B^*\to \pi$ transitions, the form factors at $q^2=0$ have been evaluated in Ref. \cite{Chang:2016cdi} using this model.
We have checked these values of form factors at $q^2=0$ and have evaluated within the same model, the form factors corresponding to $D^*\to K$ and $D^*\to \pi$ weak transitions. The results for the different form factors and the values of pole masses used in our evaluations are shown in Table \ref{table2}.
As long as the $P^*$ pole contribution to the decay amplitude is subleading, we should take the numerical contribution due to the $P^*\to P_1$ form factors as a good estimate of their true values.

\subsection{Hadronic invariant mass distributions}

\begin{figure}[h]
\includegraphics[scale=0.6]{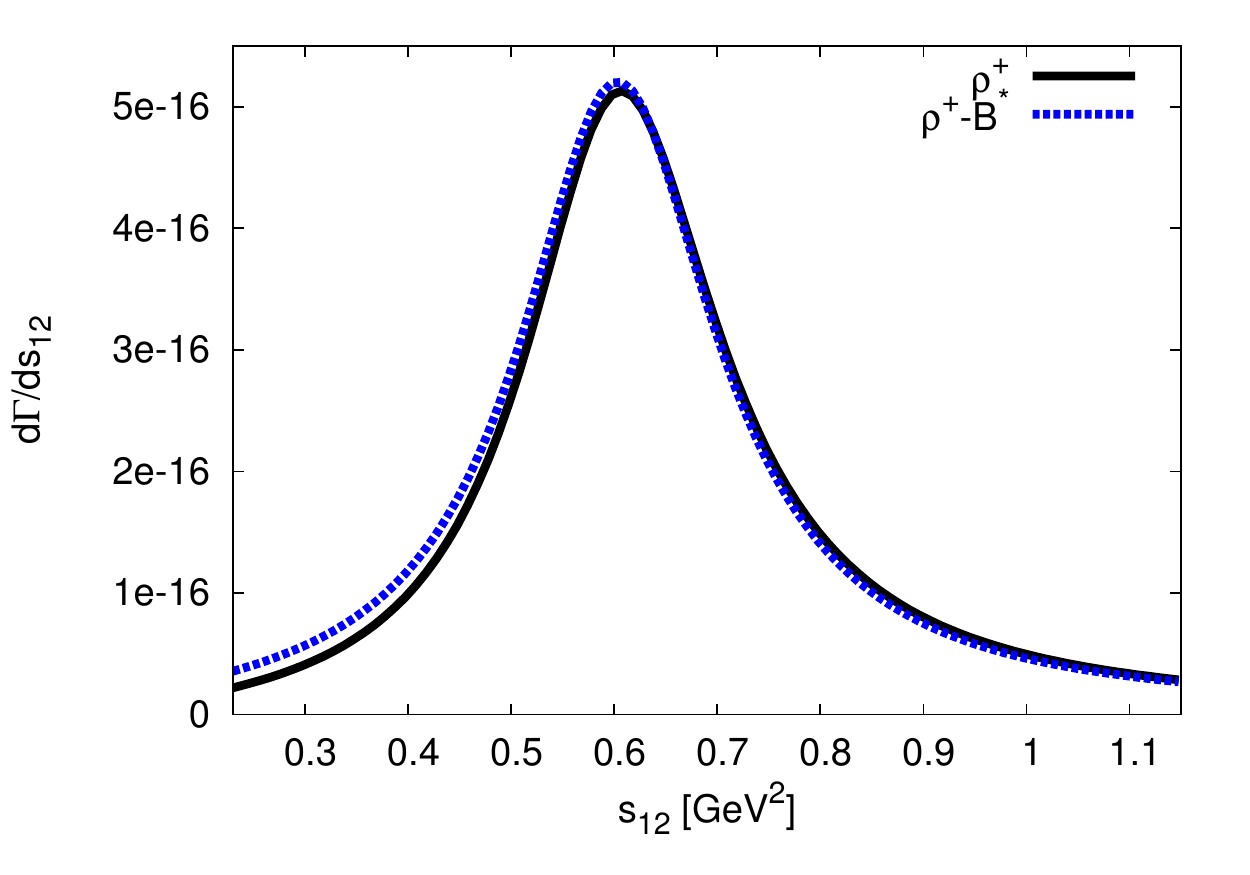}
\includegraphics[scale=0.6]{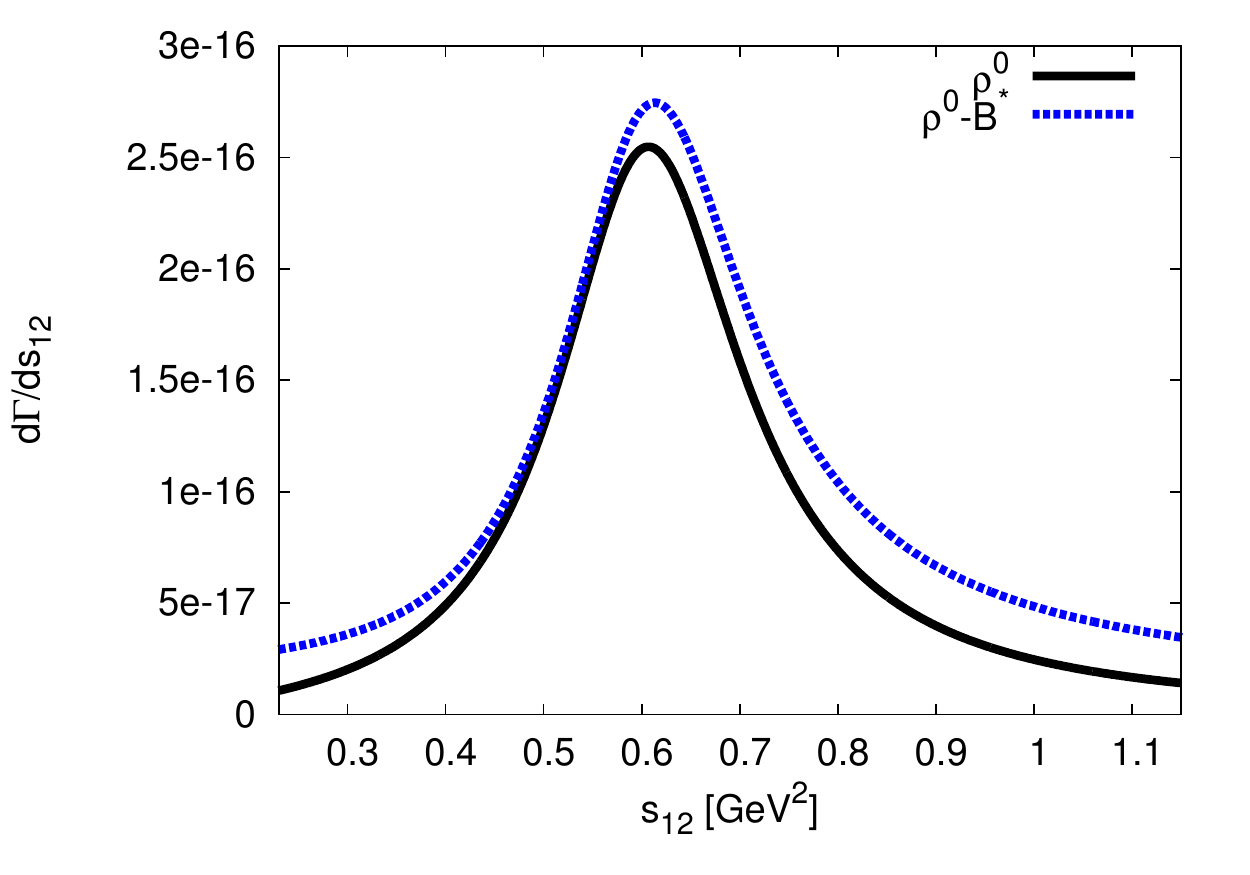}
\caption{Invariant-mass distribution of $\pi\pi$  in $B\to \pi\pi \ell^- \bar{\nu} $ ($\ell=e\mu$) decays. Left (right) panel is for decays of neutral (charged) $B$ mesons. The solid (dotted) lines describe the dominant $\rho$ ($\rho+B^*$) pole contributions to the hadronic vertex. } \label{figu2}
\end{figure}

\begin{figure}[h]
\includegraphics[scale=0.6]{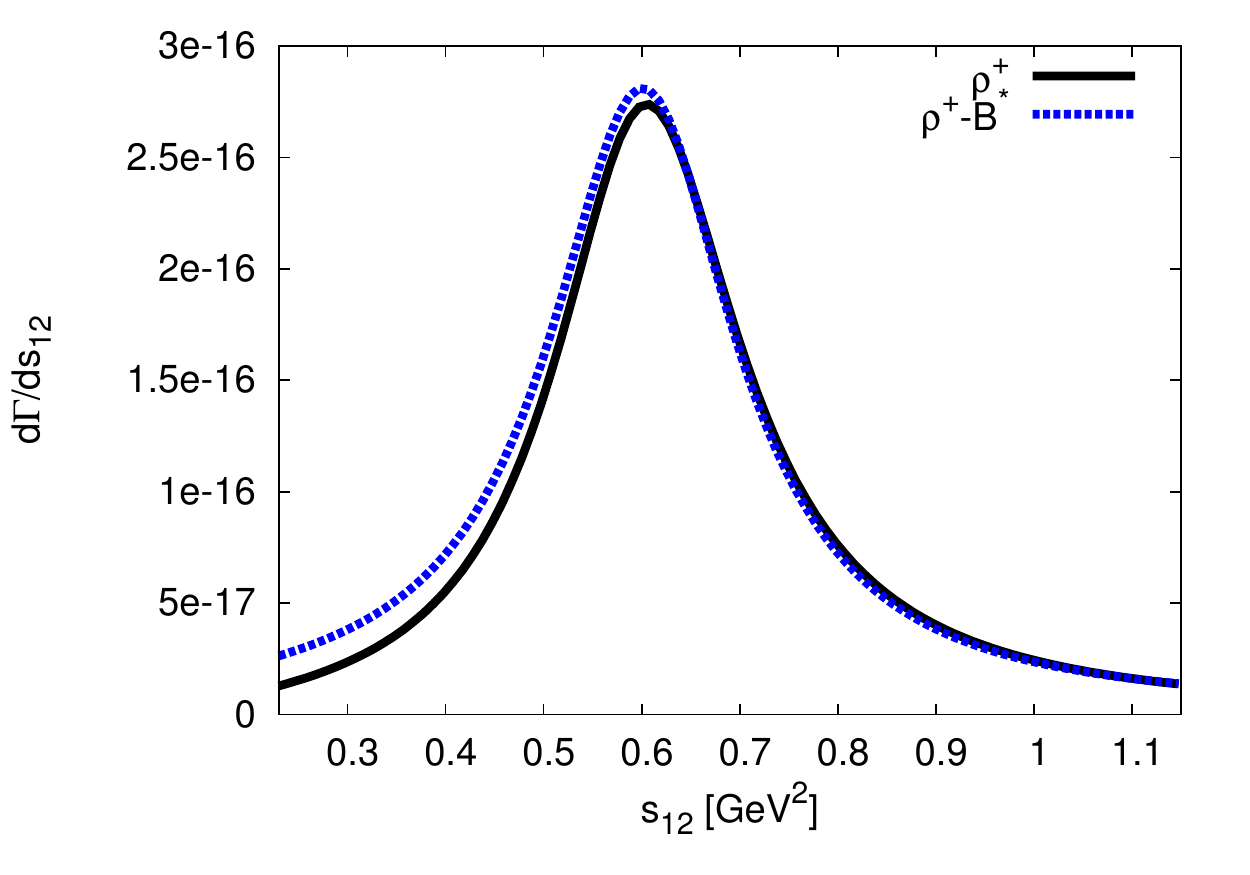}
\includegraphics[scale=0.6]{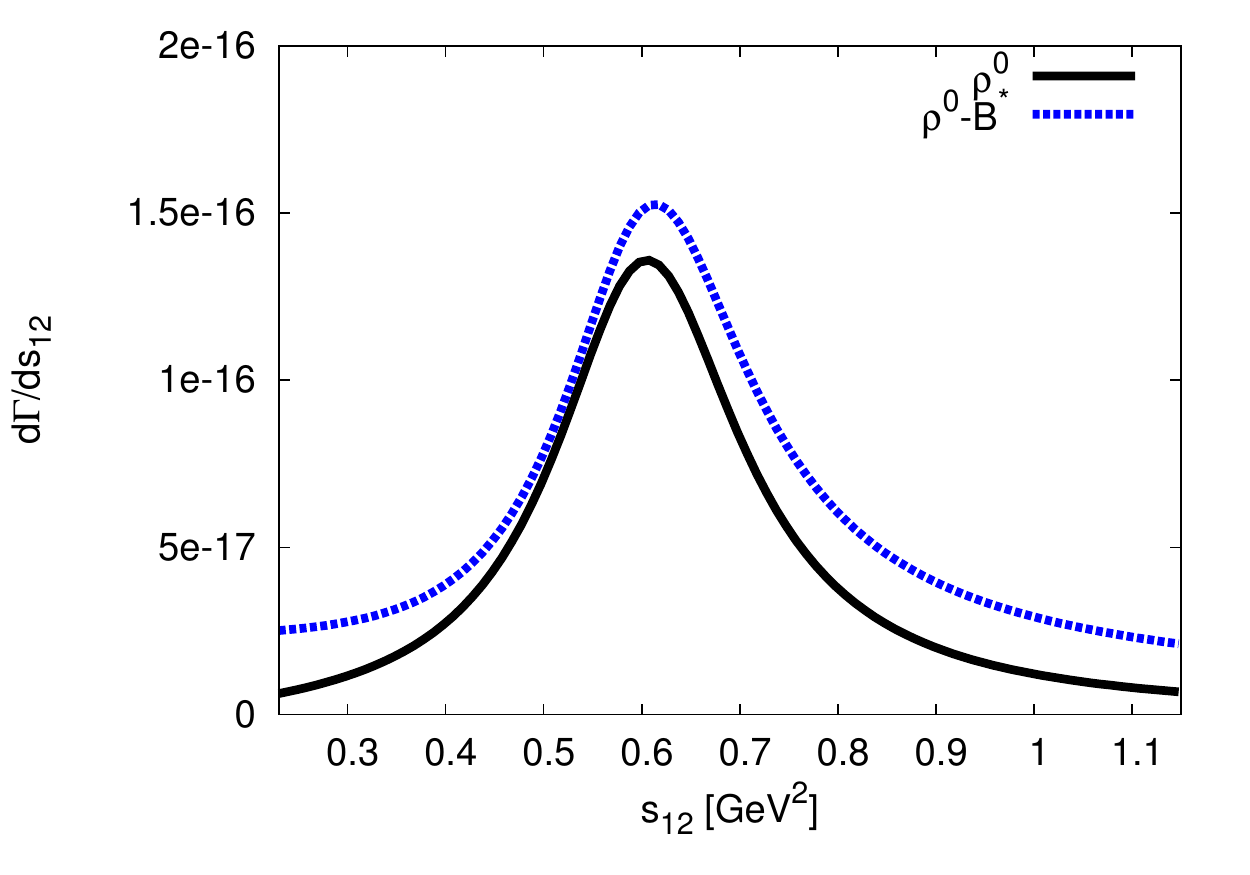}
\caption{Same as Figure \ref{figu2} for  $B\to \pi\pi \tau^- \bar{\nu} $ decays.} \label{figu3}
\end{figure}

\begin{figure}[h]
\includegraphics[scale=0.62]{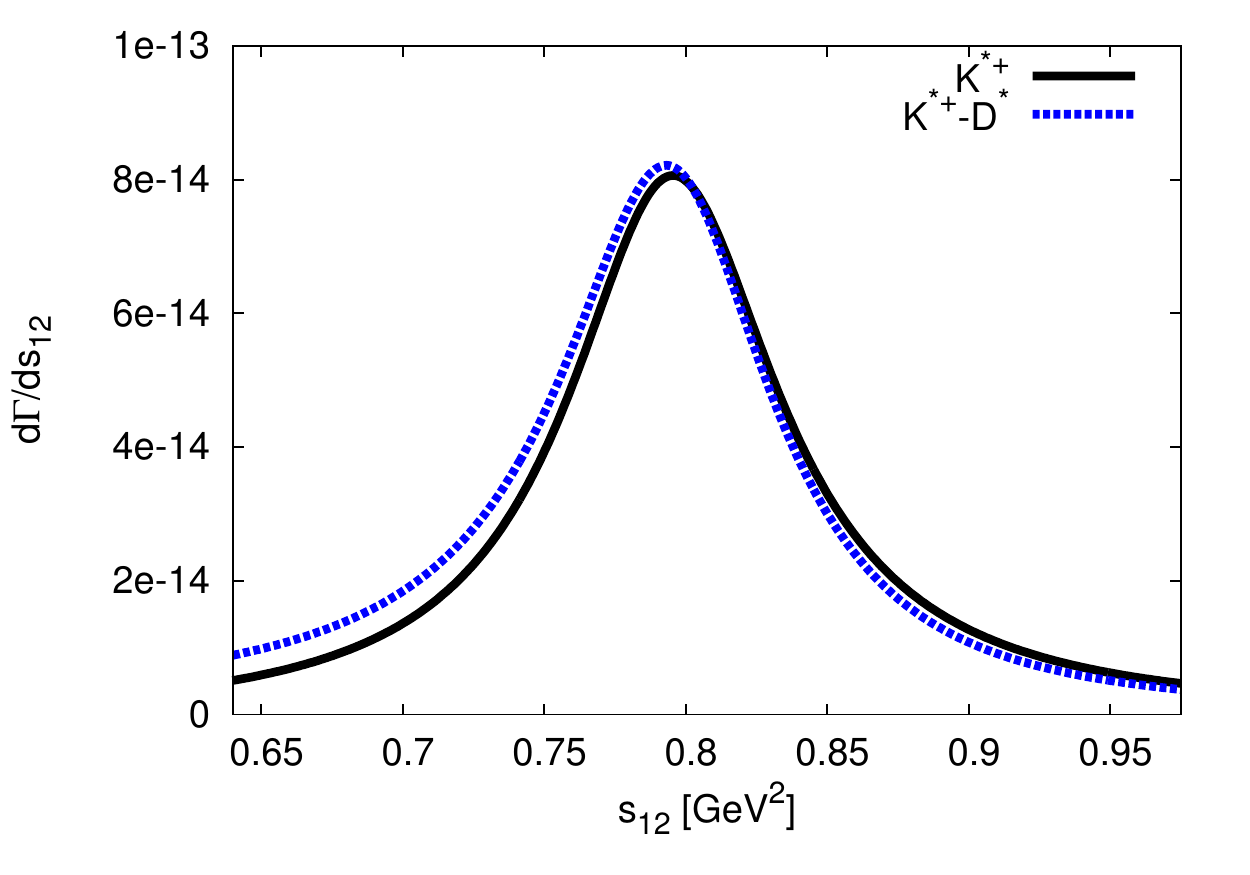}
\includegraphics[scale=0.62]{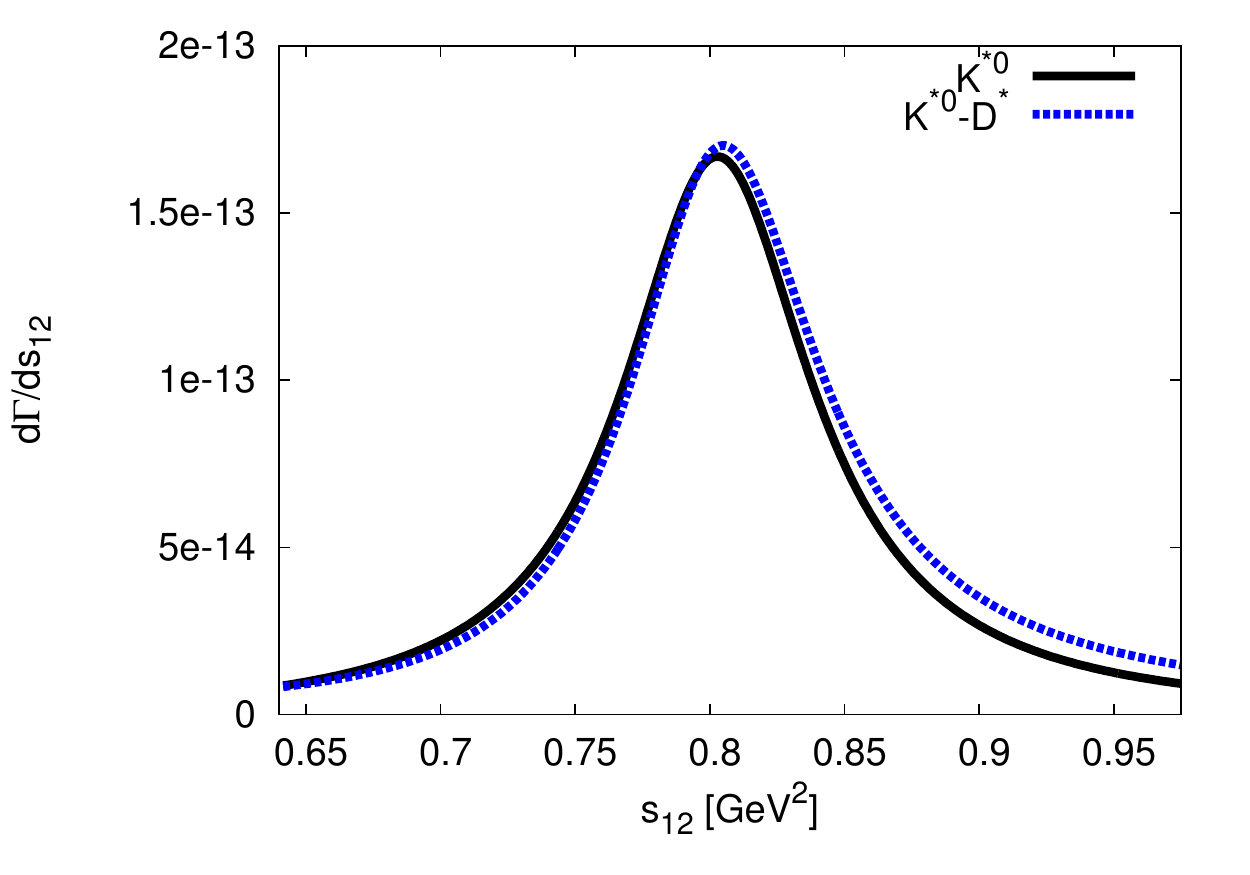}
\caption{Invariant-mass distribution of $K\pi$ in $D\to K\pi \ell^- \bar{\nu}$ decays. Left (right) panel for decays of $\bar{D}^0$ ($D^-$) meson. The solid (dotted) lines represent the $K^*(892)$ ($K^*+D^*$) pole contribution.} \label{figu4}
\end{figure}

\begin{figure}[h]
\includegraphics[scale=0.6]{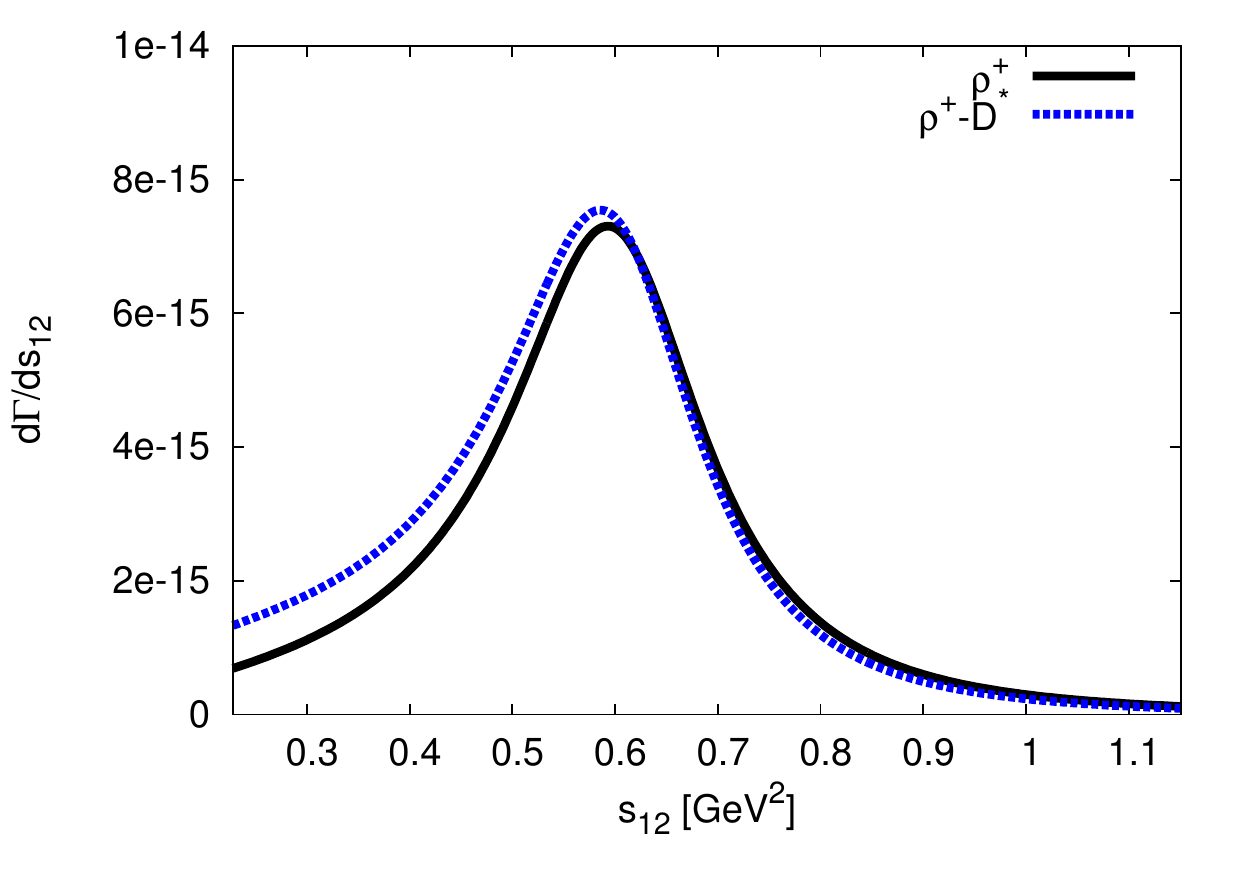}
\includegraphics[scale=0.6]{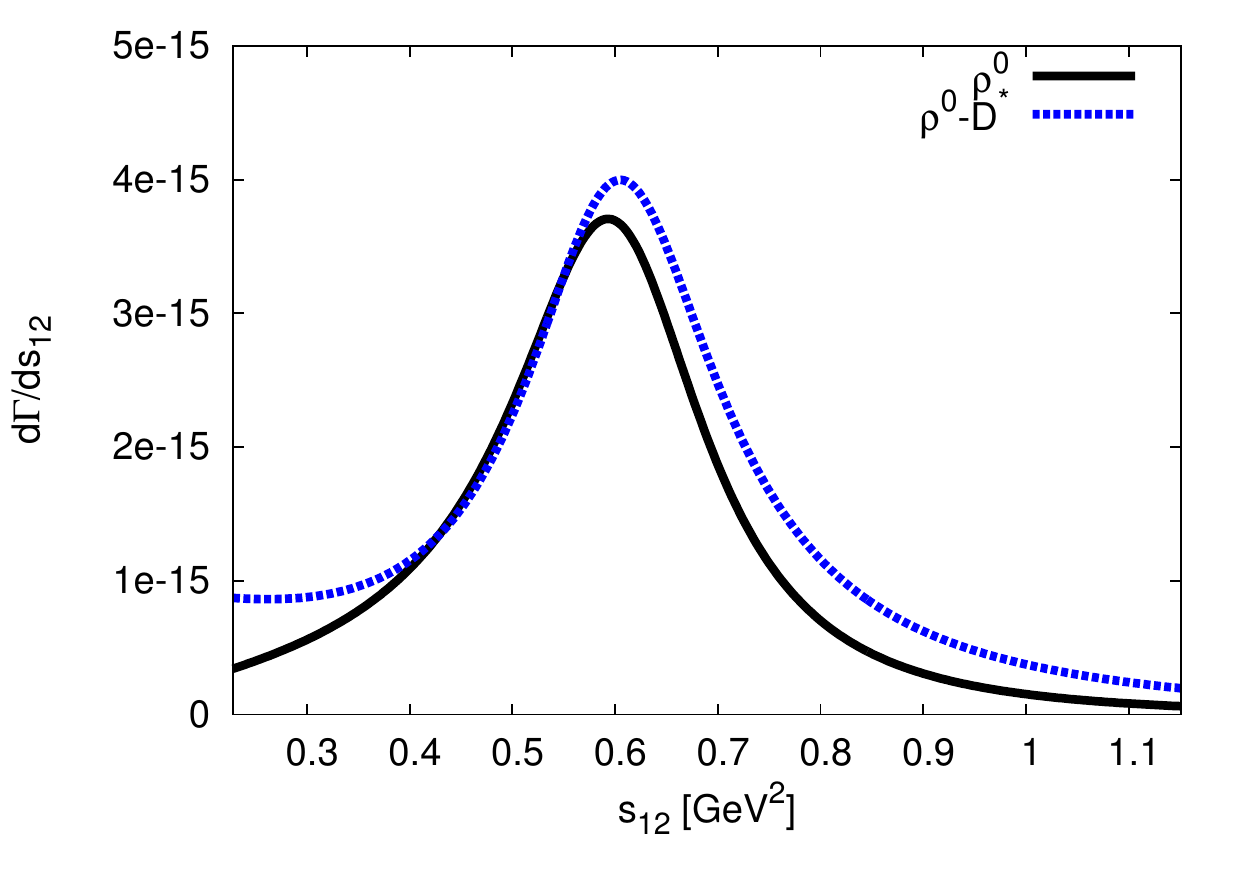}
\caption{Same as Figure \ref{figu4} for $D\to\pi\pi \ell^- \bar{\nu} $ decays.} \label{figu5}
\end{figure}

In Figures \ref{figu2}, \ref{figu3}, \ref{figu4} and \ref{figu5} we plot the hadronic invariant mass distributions of $B\to \pi\pi \ell \nu_{\ell}$ and $D\to (K\pi, \pi\pi)\ell \nu_{\ell}$ decays and compare the single dominant resonance contribution (solid line) with the full calculation including both poles (dotted line). In the case of $B$ meson decays, we have plotted separately these distributions for light and heavy $\tau$ leptons  given the interest for a test of lepton universality. We do not show the corresponding plots for $B\to D\pi \ell \nu_{\ell}$ decays because the effect of the additional pole in that case is indistinguishable.

 A comparison of the left and right panels in each of Figures \ref{figu2}-\ref{figu5}  shows important isospin breaking effects: the full contributions shift the peak of the distributions to the left (right) of the single dominant pole contribution for decays of neutral (charged) mesons. The origin of this asymmetry lies in the relative signs and different isospin factors for  couplings  of charged and neutral resonances coupled to two pseudoscalar mesons.
A fit to the $P_1P_2$ invariant mass distribution, aiming to extract the resonance parameters of the $P_{12}^*$ intermediate state in semileptonic decays, should take into account the two pole contributions. The $P^*$ pole contribution in this case, will play the role of a non-resonant background. A visual inspection of the plots in Figures \ref{figu2}-\ref{figu5} indicates that the $P^*$ pole will increase (decrease) the mass of the $P_{12}^*$ resonance when extracted from neutral (charged) heavy pseudoscalar meson decays with respect to the case where the contribution of Figure \ref{figu1} is neglected.

\subsection{Branching fractions}

  We can compute the integrated rates of $P\to P_1P_2\ell \nu_{\ell}$ decays by integrating the hadronic invariant-mass distributions as shown in Eq. (\ref{integratedrate}). We  restrict this integration to the region close to the mass of the dominant vector resonance $R\to P_1P_2$, namely $s_{12}^{\pm}=(m_{R}\pm \Gamma_{R}/2)^2$. Our results are shown in Table \ref{table1}. We can identify the resulting  decay rate\footnote{
  Of course, this is true in the case that experiments have removed the contributions of excited resonances in the $P_1P_2$ system or that they are well separated from the dominant resonance region. }
  with $\Gamma(P\to P_1P_2\ell \nu_{\ell})=\Gamma(P\to V\ell \nu_{\ell})\times B(V\to P_1P_2)$ only in the case that the contribution of Figure \ref{figu1}(a) is neglected (second column in Table \ref{table1}). When the contribution of diagram in Figure \ref{figu1}(a) is included, the correct formula necessary to extract the branching fraction of the semileptonic $P\to V$ transition is:
\be
B(P\to V\ell \nu_{\ell})= \frac{\tau_{P} \cdot \Gamma(P\to P_1P_2\ell \nu_{\ell})}{B(V\to P_1P_2)\cdot (1+\delta_{P^*})}\ , \label{factor}
\ee
where $\tau_P$ is the lifetime of the decaying particle and $\delta_{P^*}$ is the small correction due to subdominant pole contribution.

\begin{table}[b]
\centering
\begin{tabular}{|c|c|c|c|}
\hline
Channel & \ $\Gamma(P_{12})$\   & \ $\Gamma(P^*_{12}+P^*)$ \ & \ $\delta_{P^*}$\  \\
\hline \hline
$D^- \to K^+ \pi^-\ell^- \bar{\nu}_{\ell}$ & 22.2 & $22.7\pm1.2$ & 2.2 \%\\
 & (22.3) & ($22.6\pm 1.2$) & (1.3 \%) \\
\hline
$\bar D^{0} \to K^+ \pi^0\ell^- \bar{\nu}_{\ell}$ & 11.5 & $11.7\pm0.6$ & 1.7 \%\\
& (11.5) & ($11.7\pm 0.6$) & (1.7 \%)\\
\hline
$D^- \to \pi^+ \pi^-\ell^- \bar{\nu}_{\ell}$ & 1.33 &  $1.47\pm0.16$ & 10.5 \% \\
&(1.34) & ($1.45\pm 0.16$) &(8.2 \%) \\
\hline
$\bar D^{0} \to \pi^+ \pi^0\ell^- \bar{\nu}_{\ell}$ & 2.62 &  $2.69\pm0.29$ & 2.7 \% \\
& (2.64) & ($2.69\pm0.29$) & (1.9 \%) \\
\hline
$B^- \to \pi^+ \pi^-\ell^- \bar{\nu}_{\ell}$ & 9.22 &  $10.10\pm0.34$ & 9.5 \%\\
& (9.23) & ($9.86\pm 0.33$) & (6.4 \%) \\
\hline
$B^- \to \pi^+ \pi^-\tau^- \bar{\nu}_{\tau}$ & 4.92 &  $5.67\pm0.44$ & 15.2\% \\
& (4.93) & ($5.49 \pm 0.43$) & (10.2 \%) \\
\hline
$\overline{B^0} \to \pi^+ \pi^0\ell^- \bar{\nu}_{\ell}$ & 18.57 & $18.91\pm0.70$ & 1.8 \% \\
& (18.59) & ($18.91\pm 0.70$) & (1.7 \%) \\
\hline
$\overline{B^0} \to \pi^+ \pi^0\tau^- \bar{\nu}_{\tau}$ & 9.91 & $10.23\pm0.90$ & 3.2 \% \\
& (9.93) & ($10.21\pm 0.90$) & (2.7 \%) \\
\hline
$\overline{B^0}\to D^+\pi^0\ell^-\bar{\nu}_{\ell}$ & 606.2 & $606.2\pm 31.6$ &  0 \% \\
\hline
$\overline{B^0}\to D^+\pi^0\tau^-\bar{\nu}_{\tau}$ & 152.8 & $152.8\pm 8.0$ &  0 \% \\
\hline
\end{tabular}\caption{Integrated rates in units of $10^{-15}$ ($10^{-17}$) GeV for $D$ ($B$) meson decays.  The dominant (full) pole contribution is shown in the second (third) column. The quoted uncertainties arise from uncertainties in form factor inputs. Within parenthesis we have indicated the results obtained in the narrow width approximation (see text), except for the last two rows that does not change.}\label{table1}
\end{table}

The numerical value of $\delta_{P^*}$ is obtained from the ratio of the fourth/third columns in Table \ref{table1} (see last column); this correction can be as large as 15\% for $B^- \to \pi^+ \pi^-\tau^- \bar{\nu}_{\tau}$ decays. Since the effect of the additional $P^*$ pole is to increase the decay rates compared to the cases where it is neglected, the values extracted for $|V_{qq'}|$ will be decreased by $\delta_{P^*}/2$ when comparing the experimental and theoretical values of $P\to V\ell\nu$.

  For comparison, we also show in Table \ref{table1} the results obtained in the narrow width approximation for the dominant resonant contribution (figures within parenthesis). We have implemented this limit by replacing the propagator of the $P^*_{12}$ resonance as follows:
\be
\frac{1}{\left| s_{12}-m_R^2+im_R\Gamma_R \right|^2} \to \frac{\pi}{m_R\Gamma_R}\delta(s_{12}-m_R^2)\ .
\ee
Using this approximation in the integrand of Eq. (\ref{integratedrate}), the integration over the five-dimensional phase-space, reduces to an integration over  four dimensions. As it can be observed, the corresponding results change only slightly compared to the ones obtained by integrating over the finite range $(m_R-\Gamma_R/2)^2 \leq s_{12}\leq (m_R+\Gamma_R/2)^2$, except for $B^0\to \pi^+\pi^-\ell \nu_{\ell}$ decays, where the largest variations are obtained.

\section{Effects on the evaluation of CKM matrix elements}

As discussed in Refs. \cite{Kang:2013jaa, Meissner:2013pba, Albertus:2014xwa}, the value of  $|V_{ub}|$ is increased  if one uses the four-body $B\to (\pi\pi)_{\rho} \ell \nu_{\ell}$ ($B_{\ell 4}$) decays\footnote{The notation $(P_1P_2)_V$ means that the invariant mass of the pair of pseudoscalar mesons is taken in a small window around the $V$ meson mass.}, instead of the corresponding three-body $B\to \rho\ell \nu_{\ell}$ decay in its determination. This happens owing to the dynamics of the strong interactions manifested as rescattering effects and orbital angular configurations of the $\pi\pi$ system different from $L=1$. In this section we consider the additional modification of the $|V_{qq'}|$ mixing owing to strong interactions in the initial state of $P_{\ell 4}$ decays described in this paper. As a general trend, those effects tends to decrease the value of the CKM matrix element extracted from $P_{\ell 4}$ decays.

As an illustrative example let us  estimate the effect on the extraction of $|V_{ub}|$ due to the additional pole contribution in the case of charmless $B\to (\pi\pi)_{\rho} \ell \nu$ decays as measured by the Belle collaboration in Ref.  \cite{Sibidanov:2013rkk}. A rigorous procedure should include a fit to the measured $q^2=s_{34}$ distribution in order to determine the free constants of a given form factor model and then extract the value of $|V_{ub}|$ from the measured branching fraction. Instead, we estimate  the effect of the $B^*$ pole contribution  using Eq. (\ref{factor}), which is equivalent to the formula given in Ref. \cite{Sibidanov:2013rkk} in the absence of the $\delta_{P^*}$ term:
\be
|V_{ub}| =\sqrt{\frac{B(B\to (\pi\pi)_{\rho}\ell \nu_{\ell})}{\tau_B\cdot \Delta \zeta \cdot (1+\delta_{P^*})B(\rho\to \pi\pi)}} \ . \label{ckm}
\ee
Here, $B(B\to (\pi\pi)_{\rho}\ell \nu_{\ell})$ is the measured branching fraction and $\Delta \zeta=\int d\Gamma/|V_{ub}|^2$ the normalized (to the squared $|V_{ub}|^2$ quark mixing matrix element) rate integrated over the $s_{12}^{\pm}=(m_{\rho}\pm 2\Gamma_{\rho})^2$ window, and $\tau_B$ is the $B$ meson lifetime. The quantity $\Delta \zeta$ depends of the model used to describe the form factors of the $B\to \rho$ transition. Using the model of Ref. \cite{DelDebbio:1997ite}, as done by Belle in  Ref. \cite{Sibidanov:2013rkk}, which we have used also in our evaluations of the hadronic spectrum and branching fractions, we have obtained $\Delta \zeta=(13.8\pm 2.9)$ ps$^{-1}$ for the range $|\sqrt{s_{12}}-m_{\rho}|< 2\Gamma_{\rho}$.
As a check of our calculation, using the narrow width approximation and the value of $|V_{ub}|$ as in Ref. \cite{Sibidanov:2013rkk}, we reproduce the value  $\Delta \zeta=(16.5\pm 3.5)$ ps$^{-1}$ as reported in that reference for the form factor model of \cite{DelDebbio:1997ite}.

Using the branching fractions $B(B^- \to (\pi\pi)_{\rho^0}\ell^-\bar{\nu}_{\ell})=(1.83\pm 0.10\pm 0.10)\times 10^{-4}$ and $B(\bar{B}^0 \to (\pi\pi)_{\rho^+}\ell^-\bar{\nu}_{\ell})=(3.22\pm 0.27\pm 0.24)\times 10^{-4}$ as reported in \cite{Sibidanov:2013rkk} for $|\sqrt{s_{12}}-m_{\rho}|< 2\Gamma_{\rho}$, and using $\delta_{B^*}=2.8\ (24.0)\%$ for the $B^*$ pole correction in the same range of the invariant mass of $\pi^+\pi^- (\pi^+\pi^0)$ system, we get
\be
|V_{ub}|=\left\{ \begin{array}{c}
(3.87 \pm 0.46)\times 10^{-3} \ {\rm from \ } B^-\ {\rm decay} \cr
(3.62 \pm 0.40) \times 10^{-3} \ {\rm from \ } \bar{B}^0\ {\rm decay}  \cr
\end{array} \right. \ ,
\ee
where the $B^*$ pole effects mainly affects the decays of the charged $B$ meson. 
The weighted average of the above results is $|V_{ub}|=(3.73\pm 0.30)\times 10^{-3}$, which is  closer to the determination obtained from $B\to \pi \ell \nu$ decays  $|V_{ub}|=(3.72\pm 0.19)\times 10^{-3}$ as reported by the PDG \cite{Olive:2016xmw}. Let us mention that using the narrow width approximation as in Ref. \cite{Sibidanov:2013rkk}, the effect becomes smaller; using the same input data, and the corresponding values of the $B^*$ pole correction, $\delta_{B^*}=1.7\ (6.4)\%$ as shown in Table \ref{table1}, we would have obtained $|V_{ub}|=(3.57\pm 0.29)\times 10^{-3}$ for the average from $B^-$ and $\bar{B^0}$ decays.

  The effect of the additional pole considered in this paper will be also non-negligible for improved measurements of Cabibbo-suppressed $D\to \pi\pi \ell^- \bar{\nu}_{\ell}$ decays. The invariant mass distribution of the $\pi\pi$ system measured by CLEO \cite{CLEO:2011ab} for $|m_{\pi\pi}-m_{\rho}|\leq 150$ MeV is dominated by the $\rho(770)$ resonance. By assuming a monopolar form of the different form factors and assuming $|V_{cd}|=0.2252\pm 0.0007$ from the unitarity of the CKM matrix, the values $V(0)=0.84\pm0.09^{+0.05}_{-0.06}$,  $A_1(0)=0.56\pm0.01^{+0.02}_{-0.003}$ and $A_2(0)=0.47\pm0.06\pm0.04$ were derived from the measured branching fractions and invariant mass distributions \cite{CLEO:2011ab}. 

In order to estimate the effect of the $D^*$ pole contribution in the determination of $|V_{cd}|$ we can use the same form factors and branching fractions measured in \cite{CLEO:2011ab} using  Eq. (\ref{ckm}).  Since Ref. \cite{CLEO:2011ab} uses a different resonant shape of the $\pi\pi$ invariant mass than ours, for the purposes of estimating the effect of the $D^*$ pole contribution we will use our results in the narrow width approximation (in this case $\Delta \zeta=(4.0\pm 0.4)\times 10^{10}$ s$^{-1}$) and the values of $\delta_{D^*}$ are given in Table \ref{table1}. By including the $\rho$ and $D^*$ poles, we obtain
\begin{equation}
|V_{cd}|_{\rho+D^*}=0.224\pm 0.011~,
\end{equation}
from the average of $D^-$ and $\overline{D^0}$ semileptonic decays. For comparison, the value obtained by including only the $\rho$ meson resonance is $|V_{cd}|=0.230\pm 0.012$, namely the effect of including the $D^*$ pole shifts downwards the value of this CKM matrix element by 2.7\%. Although this effect is small compared to the statistical uncertainty of current measurements, it will become relevant in analyses of improved measurements in the future.

Let us emphasize that the aim of our evaluations is to estimate the shift produced by the additional pole contribution in the determination of $|V_{ub}|$ and $|V_{cd}|$ mixing matrix elements.
More refined analysis which includes the effects of the $s$-wave $\pi\pi$ system and more precise measurements of the $D(B)\to \pi\pi \ell^-\bar{\nu}_{\ell}$ branching fractions would allow to assess correctly the size of the additional $D^*(B^*)$ pole contribution. Conversely, by using precise measurements of $P_{\ell 4}$ observables combined with the most reliable determinations of the quark mixing elements would allow to test the form factor models describing the dominant $B\to V$ weak transition as well as to understand some underlying dynamics of the hadronic system.

\section{Summary and conclusions}
Precise determinations of the $b$-quark mixing matrix elements are necessary to find  possible sources of CP violation beyond the CKM mechanism. One way to reach this goal is to combine mixing values extracted from different decays of $b$-flavored hadrons. Solving current discrepancies between the most precise determinations of $|V_{qb}|$ ($q=u,\ c$) from exclusive and inclusive channels, combined with more precise calculations of form factors and refined measurements at $b$-factories will provide a consistency test of the SM and look for possible effects of new physics.
In this paper we have studied the $P\to P_1P_2 \ell \nu_{\ell}$ semileptonic decays of $B$ and $D$ mesons by modelling the weak hadronic matrix element of the $P\to P_1P_2$ transition with two poles contributions. The well known $P\to V(\to P_1P_2)\ell\nu_{\ell}$, with $V$ a resonant vector meson pole, gives the dominant contribution for invariant masses of the $P_1P_2$ within a small window  around the $V$ meson mass. A subleading tree-level additional pole contribution is identified which becomes relevant for decay observables at the few-percent level.

We have considered the effects of the subleading pole contribution in the invariant mass distribution of the meson pair in the $D\to (K\pi,\pi\pi) \ell \nu_{\ell}$ and $B\to (D\pi, \pi\pi)\ell \nu_{\ell}$ semileptonic decays. This correction shifts the invariant mass distribution differently for decays of charged and neutral heavy mesons owing mainly to different isospin factors of the strong vertex involved in each case. We have also evaluated the correction  in the branching fractions of these four-body decays induced by this subleading pole. We have illustrated how these corrections affects the determination of the $|V_{ub}|$ matrix element extracted by Belle \cite{Sibidanov:2013rkk} from $B\to \pi\pi\ell\nu_{\ell}$ decays, using a window of $\pm 2\Gamma_{\rho}$ around the $\rho$ peak in the $\pi\pi$ invariant mass. The shift in the value of $|V_{ub}|$  is not significant compared to current experimental uncertainties, althought it becomes in better agreement with the determination based on $B\to \pi \ell \nu_{\ell}$ decays. Similar considerations can be applied to the extraction of CKM matrix elements from other four-body decays of $B$ and $D$ mesons. Analysis of improved data expected in future measurements of these semileptonic decays must consider the effect of the additional pole contributions discussed in this paper.

\section*{Acknowledgements}

G.L.C. and S.L.T. are grateful to Conacyt for financial support under projects 236394, 250628 (Ciencia B\'asica) and 296 (Fronteras de la Ciencia). The work of C.S.K. was supported by the NRF grant funded by Korea government of the
MEST (No.~2016R1D1A1A02936965).

\end{document}